\documentclass[aps,prd,twocolumn,amsfonts,amssymb,amsmath,eqsecnum,nofootinbib,floatfix]{revtex4}
\usepackage[dvips]{graphicx}
\usepackage{amsmath,amssymb}
\usepackage{euler}
\usepackage{psfrag,color}

\begin{document}
\title{
Analysis of Parametric Oscillatory Instability\\
     in Signal Recycled LIGO Interferometer}
\author{
V. B. Braginsky, A. Gurkovsky, S. E. Strigin and S. P. Vyatchanin}
\affiliation{
Faculty of Physics, Moscow State University, Moscow 119992,
	Russia,\\
	e-mail: svyatchanin@phys.msu.su}
\date{\today}

\begin{abstract}
We present the analysis of  undesirable effect of parametric
oscillatory instability in signal recycled LIGO interferometer. The basis for this
effect is the excitation of the additional (Stokes) optical mode,
with frequency $\omega_1$, and the mirror elastic mode, with
frequency $\omega_m$, when optical energy stored in the main
FP cavity mode, with frequency $\omega_0$, exceeds the certain
threshold and the frequencies are related as $\omega_0\simeq \omega_1+\omega_m$.
We show that possibility of parametric instability in this interferometer is relatively small due to
stronger sensitivity to detuning. We propose to ``scan'' the frequency range where parametric instability may
take place varying the position of signal recycling mirror.

\end{abstract}
\maketitle

\section{Introduction}

The full scale terrestrial interferometric gravitational wave antennae LIGO are working now and have  sensitivity,
expressed in terms of the metric perturbation amplitude, only $2$ or $3$ times worse than the planned
level of $h\simeq 1\times 10^{-21}$ \cite{abr1,abr2} (see current sensitivity curve in \cite{LIGOsens}).
In Advanced LIGO (to be realized in approximately 2012), after the improvement of the isolation
from noises in test masses (the
mirrors of the 4 km long optical FP cavities) and increasing
the optical power circulating in the resonator up to $W\simeq
830$~kW the sensitivity is expected to reach the value of $h\simeq
1\times 10^{-22}$ \cite{amaldi,ligo2}.

In \cite{bsv} we have analyzed the undesirable effect of  parametric oscillatory instability in the Fabry-Perot (FP) cavity, which may cause a substantial decrease of the antennae sensitivity or even the
antenna malfunction. This effect  appears above the certain
threshold of the optical power $W_c$ circulating in the main mode, when the difference $\omega_0 -\omega_1$ between the
frequency $\omega_0$ of the main optical mode  and the frequency
$\omega_1$ of the idle (Stokes) mode is
close to the frequency $\omega_m$ of the mirror mechanical degree
of freedom. The coupling between these three modes arises due to
the ponderomotive pressure of light in the main and
Stokes modes and the parametric action of mechanical
oscillation on the optical modes. Above the critical value of
light power  $W_c$  the amplitude of mechanical oscillation rises
exponentially as well as optical power in the idle (Stokes)
optical mode. However, E. D'Ambrosio and W. Kells have shown \cite{ak} that
if in the same unidimensional model the anti-Stokes  mode (with
frequency $\omega_{1\, a}= \omega_0 +\omega_m$) is taken into
account, then the effect of parametric instability will be
substantially dumped or even excluded. In \cite{bsv2}, we have presented
the analysis based on the model of power recycled LIGO interferometer and demonstrated
 that anti-Stokes mode could not completely suppress the effect of parametric
oscillatory  instability. As possible ``cure'' to avoid the parametric
instability we have proposed to change the mirror shape and introduce low noise damping
\cite{bv}. D. Blair with collaborators  proposed valuable idea to heat
the test masses in order to vary curvature radii of mirrors in interferometer and hence to control detuning and decrease overlapping factor between optical and acoustic modes \cite{blair1,blair2,blair3}.
Recently, the parametric instability effect was observed in experiment \cite{Corbit}.

 It is interesting that the effect of parametric instability is important not only for large scale LIGO
 interferometer but it was observed also for micro scale whispering gallery optical resonators \cite{vahala,vahala2}.

 In this paper we propose the detail analysis of parametric instability in signal recycled Advanced LIGO
interferometer (i.e. with  additional signal recycling (SR) mirror) and show that, on the one hand, the
parametric instability in this interferometer can appear at low optical power (about several Watts) but, on
the other hand, the probability that parametric instability condition will be fulfilled is small due to small relaxation
rates of optical modes (about several Hz). We also show that by varying the position of SR mirror one can change
the frequency of anti-symmetric optical mode of  interferometer and hence to scan the range
of frequencies where  parametric instability or its precursors may arise.

In section \ref{SRInt} we derive the parametric instability conditions in signal recycled LIGO interferometer,
we discuss these results in section \ref{Discussion}. The details of calculations we present in Appendices.

\section{Signal recycled interferometer}\label{SRInt}

\begin{figure}[t]

\psfrag{FP I}{FP cavity 1}
\psfrag{FP II}{FP cavity 2}
\psfrag{PRM}{PR mirror}
\psfrag{SRM}{SR mirror}
\psfrag{Tsr}{$T_\text{sr}$}
\psfrag{Tpr}{$T_\text{pr}$}
\psfrag{T1}{$T_1=T$}
\psfrag{T2}{$T_2=T$}
\psfrag{x1}{$x_1$}
\psfrag{x2}{$x_2$}
\psfrag{y1}{$y_1$}
\psfrag{y2}{$y_2$}
\psfrag{E1}{$E_1$}
\psfrag{E1in}{$E_{1in}$}
\psfrag{E2}{$E_2$}
\psfrag{E2in}{$E_{2in}$}
\psfrag{E3}{$E_3$}
\psfrag{E4}{$E_4$}
\psfrag{E5}{$E_5$}
\psfrag{E6}{$E_6$}
\psfrag{F1}{$F_1$}
\psfrag{F1in}{$F_{1in}$}
\psfrag{F2in}{$F_{2in}$}
\psfrag{F2}{$F_2$}
\psfrag{F3}{$F_3$}
\psfrag{F4}{$F_4$}
\psfrag{F5}{$F_5$}
\psfrag{F6}{$F_6$}
\psfrag{i}{$i$}
\includegraphics[width=0.45\textwidth]{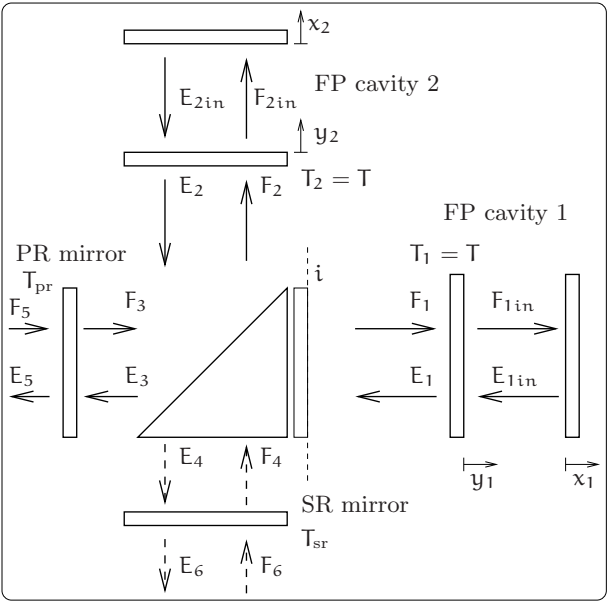}
\caption{Signal and power recycled LIGO interferometer. Here $F_1,\ E_1$
	are the field amplitudes on the input mirror of the FP cavity 1 (and on plane $(i)$), $F_2,\ E_2$
	are the field amplitudes on the input mirror of the FP cavity 2 and on the beam splitter,
	$F_3,\ E_3,\ F_4,\ E_4$  are the field amplitudes on the beam splitter.
	}\label{srscheme}
\end{figure}

We consider the LIGO interferometer with signal recycling (SR) and power recycling (PR) mirrors ---
see fig.~\ref{srscheme} and notations over there. The port $E_6$ is used for signal detection.
Interferometer is pumped via port $F_5$. Simplifications are the following:
\begin{itemize}

\item  Optical losses in all mirrors are absent (notes about generalization for non zero losses are made in the end of Sec.~\ref{Discussion}).  Suspension noise in mirrors is also  absent.

\item Transmittivity of input mirrors and length of both FP cavities  are the same.
	 They are tuned in resonance with the main mode.

\item Optical power $W$ circulating inside the arms of interferometer is constant
(approximation of constant field).
\item
	The distances between the input FP mirrors and beam splitter, and between
	the beam splitter and PR, SR mirrors are short (about several meters) --- hence we consider
	the phase advance of waves traveling between these mirrors as constant and omit
	its dependence on frequency.

\end{itemize}

\subsection{FP cavities in arms}

We denote the mean amplitude of the wave in the main mode (with frequency $\omega_0$) by cursive
capital letters and small fluctuation amplitude of the wave in the Stokes mode ($\omega_1$) by small letters.
For example, the electrical field of wave falling on back mirror  in FP cavity 1 is the following:
\begin{align*}
&E(t,\vec r_\bot)  \simeq  \sqrt\frac{2\pi}{cS_0}
        {\cal A}_{0in}(\vec r_\bot) {\cal F}_{1in}e^{-i\omega_0 t}+ \\
&+	\sqrt\frac{2\pi}{cS_1}\int_{-\infty}^\infty
        {\cal A}_{1in}(\vec r_\bot) f_{1in}(\Omega) e^{-i(\omega_1+\Omega) t}
	 \frac{d\Omega}{2\pi}+\\
&\qquad +\text{h.c.},\\
&W=   \big |{\cal F}_{1in}\big|^2, \\
& S_0= \int|{\cal A}_{0in}(\vec r_\bot)|^2\, d\vec r_\bot,\quad
 	S_1=\int|{\cal A}_{1in}(\vec r_\bot)|^2\, d\vec r_\bot.
\end{align*}
Here $W$ is the mean power in the main mode circulating inside the cavity, $c$ is the speed of light,  dimensionless functions
${\cal A}_{0in}(\vec r_\bot),\ {\cal A}_{1in}(\vec r_\bot)$ describe
the distributions of optical fields over the beam cross section, integration $\int d\vec r_\bot$ is taken over the
mirror surface. For simplicity below we consider the distributions of optical fields to be identical at all four mirrors in the arms.

We write down the displacement vector of the elastic mode with eigenfrequency $\omega_m$, for example, for the end mirror of the FP cavity 1 as a product of time and space dependent functions:
$$\left(x_1(t)\, e^{-i\omega_m t}+x_1^*(t)e^{i\omega_m t} \right)\vec u(\vec r)
$$ where $x_1$ and $x_1^*$ are  slowly varying amplitudes and $\vec u$ is spatial vector of displacements of elastic mode in the mirror.

Here we assume that the input and the end mirrors in the FP cavity 1 are elastically identical (the modes of their elastic oscillations coincide). Then we start from the formulas derived in Appendix \ref{aux} in frequency domain:
\begin{align}
\label{f1in}
f_{1in}(\Omega)&= {\cal T}_\Omega f_1 (\Omega)+ {\cal F}_{1in}\,N_1\,
	\frac{{\cal T}_\Omega\, 2ikz_1^*(\Delta -\Omega)}{i\sqrt T}\, ,\\
\label{e1}
e_1(\Omega)&= {\cal R}_\Omega f_1(\Omega) - {\cal F}_{1in}\,N_1\,
	{\cal T}_\Omega\, 2ikz_1^*(\Delta -\Omega)\, ,\\
\label{TR}
{\cal T}_\Omega &= \frac{2i\, \gamma}{\sqrt T\big(\gamma-i\Omega\big)},\quad
	{\cal R}_\Omega = \frac{\gamma+i\Omega}{\gamma-i\Omega}\, ,\\
\label{Deltaomega}
\Delta &=\omega_0-\omega_1-\omega_m,\quad k=\omega_1/c,\\
z_1(\Omega)& \equiv x_1(\Omega)-y_1(\Omega), \quad
	\gamma=cT/4L\, .\nonumber
\end{align}
Here $N_1$ is an overlapping factor (\ref{N1}), $L$ is the distance between the mirrors of the FP cavities in arms.
We also omit non-resonance terms ($\sim z_1$). In order to clarify the dependence of $z_1^*(\Delta -\Omega)$
we have to write down the last term
in (\ref{f1in}) in detail and to equate exponential quantities in the left and right parts of (\ref{f1in}):
\begin{align*}
&{\cal F}_{1in}\,e^{-i(\omega_0-\omega_1)t}\, \frac{{\cal T}_\Omega}{i\sqrt T}N_1\times
	2ikz_1^*(\Omega')\, e^{i(\omega_m+\Omega')t},\\
&-i\Omega t= -i(\omega_0-\omega_1)t + i(\omega_m +\Omega')t\ \rightarrow\
	\Omega' =\Delta  -\Omega\,.
\end{align*}

For the FP cavity 2 all formulas are the same. For the mean amplitudes we have:
${\cal E}_1 = {\cal F}_1 ,\quad  {\cal E}_2 = {\cal F}_2$.

\subsection{Beam splitter}

We consider that $F_3,\ E_3$  are the field amplitudes on the PR mirror and the beam splitter, and  $F_4,\ E_4$ are the field amplitudes on the beam splitter as it is shown in fig.~\ref{srscheme}. We assume that the beam splitter transparency is
$T_{bs}=1/2$ and the phase of the wave due to traveling between the FP cavity 2 and the
beam splitter is $e^{i\phi_2}=1$, and between the FP cavity 1 and the beam splitter is
$e^{i\phi_1}=i$. So we can imagine such plane (see Fig.~\ref{srscheme}), that phase advance
between the beam splitter and this plane is
$e^{i\phi_1}=i$ and between this plane and the input mirror of the FP cavity 1 is fold to $2\pi$.
Then $F_1,\ E_1$ are the field amplitudes on the input mirror of the FP cavity 1 (and on the plane $(i)$),
 $F_2,\ E_2$ are the field amplitudes on the input mirror of the FP cavity 2 and on the beam splitter. For the mean amplitudes we have:
\begin{align}
{\cal F}_{1}& = -{\cal F}_{3}/\sqrt 2 , \quad
	 {\cal F}_{2} = -{\cal F}_{3}/\sqrt 2 ,\quad
	{\cal E}_4= 0, \quad {\cal E}_3 = {\cal F}_3. \nonumber
\end{align}
It is convenient to introduce symmetric and anti-symmetric modes $F_\pm$:
\begin{align}
\label{f+}
F_+ &=\big(F_1+F_2\big)/\sqrt 2,\quad
	{\cal F}_+= -{\cal F}_3,\quad
	f_+= - f_3\, ,\\
\label{f-}
F_-&=\big( F_1-F_2 \big)/\sqrt 2, \quad
	{\cal F}_-=0,\quad
	f_-= - if_4\, .
\end{align}
Then for fluctuation fields $e_3$ and $e_4$ we have:
\begin{align}
\label{e4}
e_4 &= - {\cal R}(\Omega) f_4 -
	{\cal F}_{1in}\,{\cal T}_\Omega\,  \sqrt 2\,N_1 kz_-^*\, ,\\
\label{e3}
e_3 &= {\cal R}(\Omega) f_3 +
	{\cal F}_{1in}\,{\cal T}_\Omega\, \sqrt 2 \,i\, N_1k z_+^*,\\
z_-&=(x_1-y_1) - (x_2-y_2),\quad z_+=(x_1-y_1) + (x_2-y_2)\, .\nonumber
\end{align}
We see that symmetric and anti-symmetric  modes can be analyzed separately.
In subsection \ref{sym}  we consider the symmetric mode which interacts with sum coordinate $z_+$ and in subsection \ref{asym}: the anti-symmetric mode interacting with differential coordinate $z_-$.
It is worth to note that such consideration is possible only if all four mirrors in the FP cavities in arms are optically and elastically identical, just as we have assumed. For the opposite case, when the eigenfrequencies of elastic modes are different we can consider elastic mode  only in one mirror and assume other mirrors to be fixed (see subsection \ref{diffMirrors}).

\subsection{Power Recycling Mirror and symmetric mode}\label{sym}

Assuming that PR cavity is in resonance, we obtain  two equations in time domain
describing coupling between optical and elastic modes (see details of calculations in  Appendix \ref{PR}) :
\begin{align}
\label{fin+}
\dot  f_{in+}+\gamma_{0+} f_{in+}&=
	\frac{ iN_1{\cal F}_{1in}\,  \omega_1 }{\sqrt 2\, L}	z_+^*e^{-i\Delta  t},\\
\label{eqz+}
\dot z_+^* +\gamma_m  z_+^*&\simeq
	\frac{2\sqrt 2\, N_1^*{\cal F}_{1in}^*}{\pi\, i\omega_m m\, \mu}
f_{in+}(t)\,e^{i\Delta  t}\, .
\end{align}
Here $\gamma_{0+}$ is the relaxation rate of symmetric mode (\ref{gamma0+}), $\gamma_m$  is the relaxation rate of elastic mode, $m$ is the mass of each of the mirrors, and $\mu$  is normalizing constant (\ref{mu}).

We find the solutions of  this set of equations
(\ref{fin+}, \ref{eqz+}) in the form:
\begin{align*}
f_{in+}(t)=f_{in+}e^{(\lambda-i\Delta) t }, \quad z^*_+(t)=z^*_+e^{\lambda t}\, ,
\end{align*}
and as a result we obtain the characteristic equation:
\begin{align}
\label{CharEq+}
2{\cal Q} &= (\lambda - i\Delta + \gamma_{0+})(\lambda+\gamma_m),\\
{\cal Q} &\equiv \frac{ \Lambda_1W  \omega_1 }{ c L\omega_m m}, \quad
	W= |{\cal F}_{1in}|^2 ,\\
\label{Lambda1}
\Lambda_1 &\equiv\frac{|N_1|^2}{\mu}=
	 \dfrac{V\left|\int{\cal A}_{0in}{\cal A}_{1in}^*u_\bot\, d\vec r_\bot\right|^2}{
	 \int|{\cal A}_{0in}|^2\, d\vec r_\bot \int|{\cal A}_{1in}|^2\, d\vec r_\bot
	 \int |\vec u(\vec r)|^2 \, d\vec r}.
\end{align}
Here $\Lambda_1$ is general overlapping factor between the main, Stokes optical modes, and elastic mode, $u_\bot$ is the component of displacement vector $\vec u$ of elastic mode normal to the mirror surface, $\int d\vec r_\bot$ corresponds to the integration over the mirror surface and $\int d\vec r$: to the integration over the mirror volume $V$.

We present $\lambda$ as a sum of real and imaginary parts: $\lambda=a+ib$,
substitute it  into (\ref{CharEq+}) and obtain two equations:
\begin{align}
\label{ChEq1+}
&a^2-b^2 +a\big(\gamma_m+\gamma_{0+}\big) +
	b\Delta  +\gamma_m\gamma_{0+} - 2{\cal Q}=0\, ,\\
\label{ChEq2+}
&2ab +b\big(\gamma_m+\gamma_{0+}\big) -a \Delta  -\gamma_m\Delta =0\, .
\end{align}
The parametric instability condition corresponds to $a>0$. Putting $a=0$  we find
the condition of boundary situation (between stability and instability). The additional
analysis gives the sign of inequality in the parametric instability condition.
We find formal solution for $b$ from (\ref{ChEq2+}) with assumption $a=0$, substitute it into (\ref{ChEq1+}), and obtain the parametric instability condition:
\begin{align}
\label{InstCond+}
\frac{2\cal Q}{\gamma_m\gamma_{0+}}&> 1 +\frac{\Delta^2}{\big(\gamma_m+\gamma_{0+}\big)^2}\, .
\end{align}

This condition can be compared with the condition of parametric instability for single FP cavity \cite{bsv}
with relaxation rate $\gamma$
\begin{align}
\label{RQ}
{\cal R}_0 & > 1+\frac{\Delta^2}{\gamma^2},\quad
{\cal R}_0  = \frac{\cal Q}{\gamma_m\gamma}\, .
\end{align}

We see that conditions (\ref{InstCond+}) and (\ref{RQ}) approximately coincide with each other if  one
substitutes   $\gamma_{0+}$ instead of  $\gamma$  and takes  into account  inequality
$\gamma_m\ll \gamma_{0+}$ (see estimates in Appendix~\ref{param}).  The factor $2$ appears due to
the fact that in derivation (\ref{InstCond+}) we take into account the displacements of $4$ mirrors in the interferometer while in (\ref{RQ}) the displacement of only one mirror of FP cavity is taken into account.

\subsection{Signal recycling mirror and anti-symmetric mode}\label{asym}

We assume that in general case the SR cavity is not in resonance and phase advance $\phi$
of the wave between the SR mirror and the beam splitter is an arbitrary one. We also
assume that $\phi$ does not depend on frequency $\Omega$ due to the short length of the SR
cavity.  After the calculations presented in Appendix \ref{SR} one can obtain  two equations describing the coupling between optical and elastic modes:
\begin{align}
\label{fin-}
\dot f_{in-}+\left( \gamma_{0-}-i\delta\right)f_{in-}=
	\frac{N_1{\cal F}_{1in}\,i\, \omega_1z^*_-}{L\sqrt 2}\, e^{i\Delta t},\\
\label{eqz-}
\dot z_-^* +\gamma_m  z_-^*\simeq
	\frac{2\sqrt 2\, N_1^*{\cal F}_{1in}^*}{\pi\, i\omega_m m\mu } \, f_{in-}(t)\,e^{i\Delta  t}.
\end{align}
Here $\gamma_{0-}$ is the relaxation rate of anti-symmetric mode (\ref{gamma0-}), $\delta$ is the detuning depending on the SR mirror position (\ref{delta}).

Searching for the solution of the equations set (\ref{fin-}, \ref{eqz-}) in the form:
\begin{align*}
f_{in-}(t)=f_{in-}e^{(\lambda-i\Delta) t }, \quad z_-^*(t)=z_-^*e^{\lambda t},
\end{align*}
we obtain the characteristic equation:
\begin{align}
\label{CharEq-}
2{\cal Q} &= \big(\lambda+\gamma_{0-}-i(\delta +\Delta)\big)(\lambda+\gamma_m).
\end{align}

We see that the
characteristic equation (\ref{CharEq-}) differs from the analogous one for the symmetric mode (\ref{CharEq+})
only by replacement of $ \Delta \to \ \Delta +\delta$. Hence we can write down the
condition of instability for the anti-symmetric mode using (\ref{InstCond+}):
\begin{align}
\label{InstCond-}
\frac{2\cal Q}{\gamma_m\gamma_{0-}} > 1
	+\frac{(\Delta+\delta)^2}{ \big(\gamma_m+\gamma_{0-}\big)^2}\, .
\end{align}

\begin{figure}
\psfrag{y1}{}
\psfrag{x1}{$\phi$}
\psfrag{gamma}[lb][lt]{$\gamma_{0-}/(T_{sr}\gamma)$}
\psfrag{delta}[lt][lc]{$\delta/\gamma$}
\includegraphics[width=0.45\textwidth, height=0.25\textwidth]{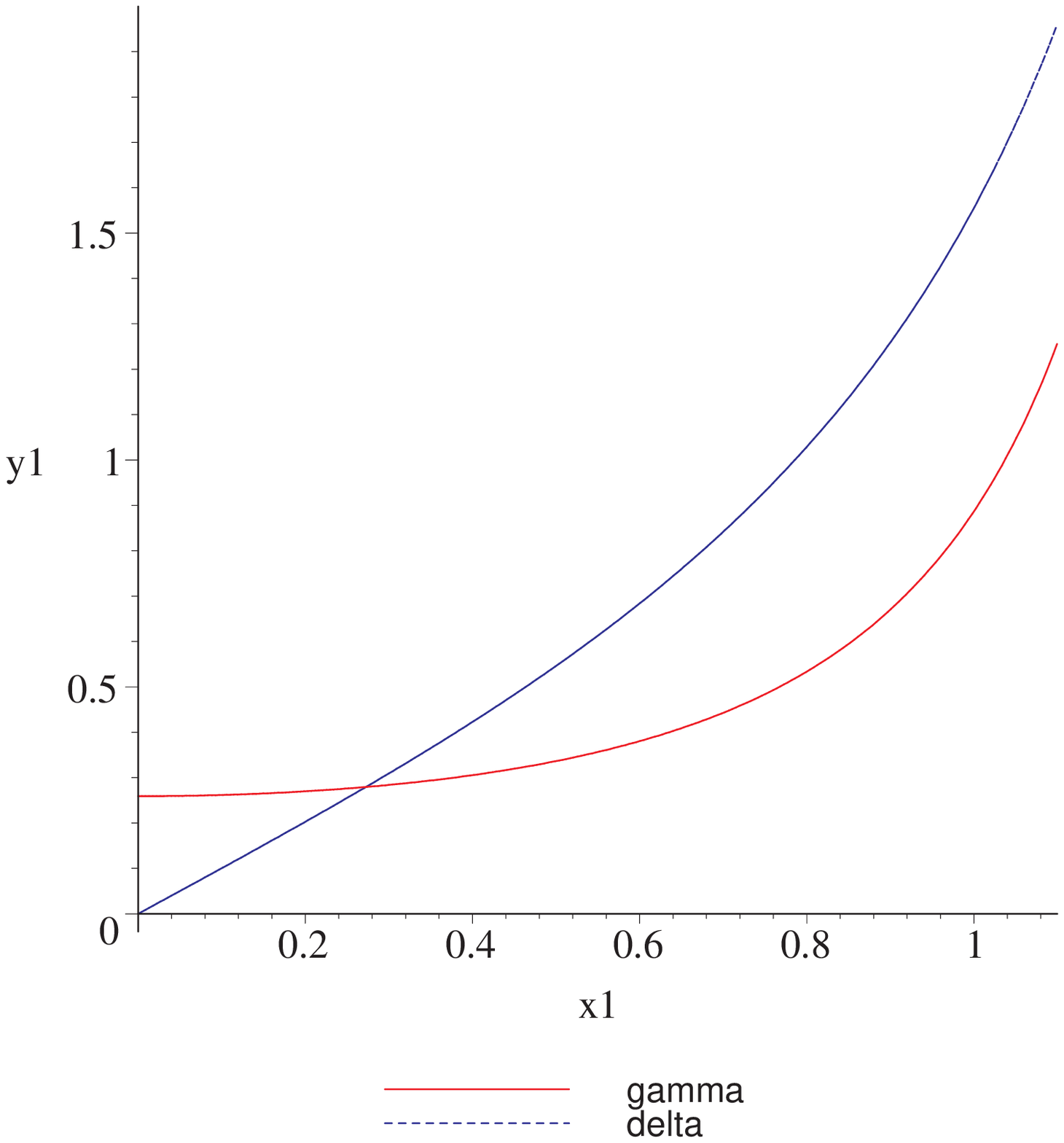}
\psfrag{y1}{}
\psfrag{x1}[rc][lc]{$\phi$}
\psfrag{gamma}[lb][lt]{$\gamma_{0-}/\gamma$}
\psfrag{delta}[lt][lc]{$\delta/\gamma$}
\includegraphics[width=0.45\textwidth, height=0.25\textwidth]{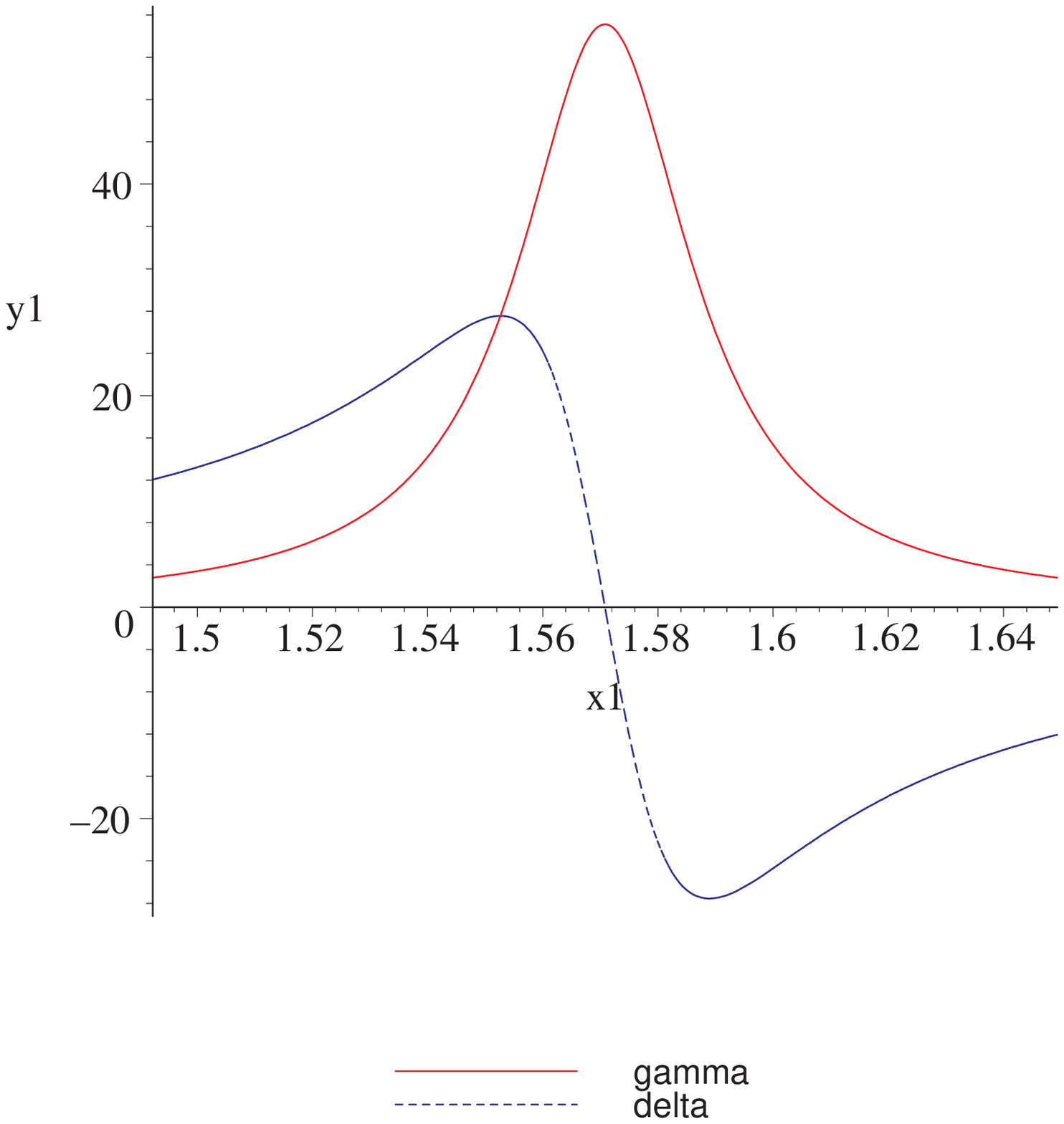}
\caption{Dependence of the relaxation rate $\gamma_{0-}$ and detuning $\delta$ of anti-symmetric mode on angle $\phi$ for $T_{sr}=0.07$ (planned in Advanced LIGO). Top: close to resonance case. Bottom: close to anti-resonance case. }\label{gd}
\end{figure}


The relaxation rate $\gamma_{0-}$ and detuning $\delta$ depend on angle $\phi$ (the position of the SR mirror). Analyzing definitions (\ref{gamma0-}, \ref{delta}) we see that there are  two cases: close to resonance ($\phi$ close to 0) and to anti-resonance ($\phi$ is close to $\pi/2$) cases. Expanding the denominator in (\ref{gamma0-}, \ref{delta}) in series over $T_{sr}$ (recall $T_{sr}\ll 1$), we obtain useful formulas:
\begin{align}
\delta &\simeq \frac{\gamma\, \sin 2\phi}
	{\dfrac{T_{sr}^2}{8}+\cos^2\phi\left(2-T_{sr}-\dfrac{T_{sr}^2}{4}\right)},\\
\gamma_{0-} &\simeq \frac{\gamma\, T_{sr} }
	{\dfrac{T_{sr}^2}{4}+2\cos^2\phi\left(2-T_{sr}-\dfrac{T_{sr}^2}{4}\right)}\, .
\end{align}

Manipulating the angle $\phi$ (via the SR mirror position variation) we can vary the relaxation rate $\gamma_{0-}$ and detuning $\delta$ in wide ranges:
\begin{align}
\label{gammarange}
\frac{T_{sr}\gamma}{4}\, \le\ & \gamma_{0-}\ \le\ \frac{4\gamma}{T_{sr}},\qquad
\frac{ - 2\gamma}{T_{sr}}\ \le\ \delta\ \le \ \frac{2\gamma}{T_{sr}}\, .
\end{align}
It is demonstrated in Fig.~\ref{gd}. (Note that the above formula is valid only while $\gamma_{0-}\ll c/L$, \textit{i.e.} $T/8T_{sr}\ll1$.)
Using formula (\ref{gammarange}), we have the following estimates for parameters of Advanced LIGO (Appendix~\ref{param}):
\begin{align*}
1.6\ {\rm sec}^{-1} &\le \gamma_{0-} \le 6\times 10^3\ {\rm sec}^{-1}  ,\\
-3\times 10^3\ {\rm sec}^{-1}& \le  \delta  \le  3\times 10^3\ {\rm sec}^{-1}.
\end{align*}
So we can ``scan'' the frequency range to find instability (or its precursors) by variation of the SR mirror position.
As precursors we may register Stokes modes providing information about the resonance frequencies of
elastic modes which ``suit'' each other by spatial distributions. These Stokes modes may be the modes of
higher orders (dipole, quadrupole and so on). In order to extract them one has to detune the output mode cleaner
which is planned to be placed after the SR mirror which is an additional but not complicated
 operation\footnote{Advanced LIGO currently plans a rigid short (tens of cm) output mode
 cleaner which can be put on resonance for the carrier but which rejects all  frequencies out of mode cleaner
 bandwidth.  Detuning, for example, to dipole mode requires change  of mode cleaner length $\ell$ by
 relatively small value
 $\delta\ell/\ell= (\omega_\text{dipole}-\omega_\text{main})/\omega_\text{main}\simeq 10^{-7}$.}.
 It provides us {\em in situ} with very valuable information about the possible danger of parametric instability.

\subsection{Different mirrors: only one mirror is in resonance}\label{diffMirrors}

Here we consider the case  when the frequencies of elastic modes in different mirrors do not coincide with each other and thus, we assume that the frequency of elastic mode of only one mirror is in resonance, for example, the end mirror in the FP cavity 1. Therefore we consider  other mirrors as fixed, and only the coordinate $x_1$ will be taken into account. Then  Eqs. (\ref{fin+}, \ref{fin-}) will be valid with substitution $x_1\ \to z_+,\ z_-$. For calculation of ponderomotive force we consider the field in the FP cavity 1 as a sum of symmetric and anti-symmetric modes fields. After calculations presented in Appendix \ref{PRSR} we get the characteristic equation in the compact form:
\begin{align}
\label{CharEqx1}
\big(\lambda+i\Delta +\gamma_m\big)=&
		\frac{\cal Q}{2}\left(\frac{1}{\lambda +\gamma_{0+} }+
			\frac{1}{\lambda-i\delta +\gamma_{0-} } \right).
\end{align}

\paragraph*{Instability condition in pure power recycled LIGO interferometer.}
Using characteristic equation (\ref{CharEqx1}) we consider the case when there is no SR mirror
in the interferometer, then we have to substitute $T_{sr}=1$ into (\ref{delta},\ref{gamma0-})
and hence $\delta=0,\ \gamma_{0-}=\gamma$ ($\gamma$ is the relaxation rate of a single
FP cavity in arm). For such power recycled configuration we have the following inequalities:
\begin{align}
\label{PRuneq}
\gamma_m \ll \gamma_{0+} \ll \gamma\, .
\end{align}
The analysis presented in Appendix \ref{PRSR} gives the following parametric instability condition:
\begin{align}
\label{PCondgamma}
		\frac{\cal Q}{2\gamma_m }\left(
		\frac{ \gamma_{0+}}{ \gamma_{0+}^2+\Delta^2} +
		\frac{ \gamma}{ \gamma^2+\Delta^2}\right)&\ge 1\, .
\end{align}
It is useful to rewrite approximation of this instability condition in particular cases for
different detunings $\Delta$:
\begin{subequations}
\label{PIPart}
\begin{align}
\Delta\ll \gamma_{0+}\ :\qquad &\frac{\cal Q}{2\gamma_m\gamma_{0+} }\ge 1,\\
\gamma_{0+}\ll \Delta \ll \gamma \ :\qquad &\frac{\cal Q}{2\gamma_m \gamma}\left(
		\frac{ \gamma_{0+}\gamma}{\Delta^2} +1 \right)\ge 1\, ,\\
\Delta \gg \gamma\ :\qquad & \frac{{\cal Q}\, \gamma}{2\gamma_m \Delta^2}\ge 1\, .
\end{align}
\end{subequations}

Note that condition (\ref{PCondgamma}) {\em slightly differs} from the formula (6) in our paper
\cite{bsv2}\footnote{Our notations relate to the notations in  \cite{bsv2} as follows:
$\gamma \Rightarrow \delta_1,\ \gamma_{0+}\Rightarrow \delta_{pr}, \ \gamma_m \Rightarrow \delta_m$,
and  parameters $\cal Q$ and ${\cal R}_0$ relate as (\ref{RQ}).}. However, the error made
in \cite{bsv2} is insignificant: the particular cases (\ref{PIPart}) {\em coincide} with analogous
particular cases obtained from formula (6) in \cite{bsv2}.

\paragraph*{The instability condition for signal recycled LIGO interferometer.}
Using estimates of Appendix~\ref{param} we have the inequality
\begin{equation}
\label{SRPRuneq}
\gamma_m \ll \gamma_{0+}< \gamma_{0-}\, .
\end{equation}
From the characteristic equation (\ref{CharEqx1}) and inequality (\ref{SRPRuneq})
we obtain the parametric instability condition:
\begin{align}
\label{SRPIcond}
\frac{\cal Q}{2\gamma_m}\left(\frac{\gamma_{0+}}{\gamma_{0+}^2 +\Delta^2} +
	\frac{\gamma_{0-}}{\gamma_{0-}^2+(\Delta+\delta)^2}\right) > \, 1\, .
\end{align}
See details of calculations in Appendix~\ref{PRSR}.

\section{Discussion}\label{Discussion}

For our discussion we use the following scale of relaxations rates (see Appendix~\ref{param}):
the relaxation rate of elastic mode $\gamma_m \simeq 6\times \left(10^{-4}\, \div \ 10^{-2}\right)\ \text{sec}^{-1}$,
the relaxation rates of symmetric and anti-symmetric modes $\gamma_{0+}\simeq 1.5\, \text{sec}^{-1}$,
 $\gamma_{0-}\ge 2\, \text{sec}^{-1}$, and the relaxation rate of a single FP cavity in arm $\gamma\simeq 100\,   \text{sec}^{-1}$.

First of all we see that the main difference between the parametric instability in a signal recycled Advanced LIGO
interferometer and in pure power recycled interferometer is the crucial dependence on detuning, compare Eqs.
(\ref{InstCond+},\ref{InstCond-},\ref{SRPIcond}) with (\ref{PCondgamma}). In power recycled interferometer the parametric instability takes place  if $|\Delta|<\gamma\simeq 100\ \text{sec}^{-1}$ while in signal recycled interferometer: $|\Delta | <\gamma_{0+}, \ \gamma_{0-}\simeq 2\ \text{sec}^{-1}$.

On the one hand, in case of relatively small detuning the parametric instability in a signal recycled interferometer takes place
at relatively low value of optical power. For example,  if $\Delta\ll \gamma_{0+}$ and $\delta \gg \gamma_{0-}$
one can obtain from  Eq.~(\ref{SRPIcond}) that the parametric  instability will take place at power
$W_c\simeq 5$~W (!) circulating in arms  (if $\omega_m =10^5\   \text{sec}^{-1}$,
$\gamma_m=6\times 10^{-4} \,\text{sec}^{-1}$, $\Lambda_1\simeq 1$). On the
other hand, there is a small chance that such small detuning
will take place and for large detuning ($|\Delta|>\gamma_{0+}$) the realization of parametric instability requires
dramatically larger optical power: $W_c \sim \Delta^2/\gamma_{0+}^2$. For example, if detuning is
about 1~kHz ($|\Delta|\simeq 6\times 10^3$ sec$^{-1}$) and other parameters are the same one can obtain
$W_c \simeq 10^8$~W (!). For the same reason the possibility that the presence of anti-Stokes mode can
depress the parametric instability is small enough especially for such detunings. Therefore, we did not consider  the
anti-Stokes mode in our analysis (``anti-Stokes'' generalization can be done using the same approach as in \cite{bsv2}).

Another factor that can decrease the possibility of parametric instability is the small value of overlapping factor; even if the
 frequencies of Stokes and elastic modes ``suit'' each other their spatial distributions (at mirrors surface)
may considerably decrease the overlapping factor $\Lambda_1$ and, hence, the possibility of parametric instability.

The elastic modes can be calculated numerically.
Unfortunately,  the numerical calculations of elastic modes of cylinder mirrors face with obvious difficulty:
the accuracy of elastic mode frequencies calculations is dramaticaly insufficient. The standard packages
FEMLAB or ANSYS,  used for this purpose \cite{bsv2,blair1}, provide the accuracy about
{\em several percents} only, while  we need the accuracy at least
$\gamma_{0+}/\omega_m\simeq 10^{-7}\, \div\, 10^{-5}$ (!). Nevertheless, numerical calculations
have sense for estimates of overlapping factors and as information (about the frequency range) for the
experimentalist where the parametric instability may occur.

Even if the numerical calculation methods will be improved to the extent that they will achieve the required accuracy, they can not solve
the  problem completely because: (i) the shape of mirrors differs from the cylinder shape, for example,
the pins to attach suspension
fiber may produce the shift of elastic mode frequency up to $100$~sec$^{-1}$ \cite{bsv2} and (ii) the inhomogeneity
of fused silica (the material mirrors should be manufactured of) may also provide an uncontrollable  shift
of elastic mode frequency \cite{bsv2}. That is why we present the separate
consideration of the case when the elastic mode of only one mirror is taken into account in subsection~\ref{diffMirrors}.

Parametric instability can also be investigated  in GEO600 configuration interferometer which also has
signal and power recycling mirrors, but has no FP cavities in arms. Hence the formulas above can be generalized
for GEO600 configuration by putting the transmittance of input mirrors in arms to be equal to $1$ and, hence, $\gamma =c/2L$.
Then the relaxation rate of symmetric mode will be about
$\gamma_{0+}^\text{GEO}\simeq T_{pr}c/4L\simeq 0.75\times 10^{-5}$~sec$^{-1}$ (we assume
$T_{pr}=0.012,\ L=1.2$~km) and using formula (\ref{InstCond+}) we estimate that parametric instability may
take place at relatively small optical power $W_c^\text{GEO}\simeq 100$~W circulating in arms if we
assume zero detuning $\Delta=0$, $m=10$~kg, $\omega_m=10^4$~sec$^{-1}$,
 $\gamma_m=10^{-4}$~sec$^{-1}$ and $\Lambda_1=1$.

Our consideration can be generalized for the case of mirrors with losses. Let each mirrors in the FP cavity to have
loss coefficient $A$ which is small compared to transparency: $\eta\equiv 2A/T< 1$. In this case
all  formulas (\ref{InstCond+}, \ref{InstCond-}, \ref{PCondgamma}, \ref{SRPIcond}) for parametric
instability conditions are valid with the following substitutions:
\begin{align}
\gamma &\Rightarrow \gamma(1+\eta), \quad
	\gamma_{0\pm}\Rightarrow \gamma_{0\pm} +\eta\gamma.
\end{align}
For Advanced LIGO the losses in mirrors are small enough (it is planned that $A\simeq 5$~ppm,
$\eta\simeq 2\times 10^{-3}$).

However, the diffractional losses may be large for optical modes with
high indices \cite{bsv2}. Kip Thorne \cite{thorne,stv} pointed out that the case of large  diffractional
losses  (when round trip relative losses are close to unity) requires a separate analysis.
This work is in progress now \cite{st}.

\section*{Conclusion}

We have shown that in the signal recycled Advanced LIGO interferometer the possibility of falling into the trap
of  the parametric instability is smaller than for the pure power recycled one due to stronger sensitivity to detuning.

We think that the most reliable method to avoid the parametric instability is the {\em direct experimental} test.
For signal recycled interferometer we have good method to investigate the possibility of
parametric instability experimentally {\em in situ}: varying the SR mirror position, one can detune the frequency of
anti-symmetric mode in wide range to find the instability or its precursors as it was shown in
subsection~\ref{asym}. It is important that for GEO600 configuration  we can introduce detuning larger than in
Advanced LIGO: the scanning may be performed inside the free spectral range $c/2L$.
This scanning combined with the detailed knowledge about the elastic modes (it can be obtained
{\em in situ} in separate experiments before the test masses are placed into the interferometer) will provide us with very valuable information helpful for avoiding the parametric instability.

We hope that parametric instability effect can be eliminated in Advance LIGO interferometer after the detailed
experimental investigations supported by theoretical analysis.

\acknowledgments

We are grateful to Farid Khalili,  David Ottaway, David Shoemaker, Ken Strain, Beno Willke, Bill Kells and  Chunnong Zhao  for valuble notes.
This work was supported by LIGO team from Caltech and in part by NSF and Caltech grant
PHY-0353775, by the Russian Agency of Industry and Science, contracts
No. 5178.2006.2 and 02.445.11.7423.

\appendix

\section{FP cavity with two movable mirrors}\label{aux}

\begin{figure}[h]
\psfrag{f1}{$f_1$}
\psfrag{e1}{$e_1$}
\psfrag{f1in}{$f_{1in}$}
\psfrag{e1in}{$e_{1in}$}
\psfrag{ft}{$\tilde f_{1in}$}
\psfrag{et}{$\tilde e_{1in}$}
\psfrag{x}{$x$}
\psfrag{y}{$y$}
\psfrag{T}{$T,\quad m_1$}
\psfrag{r}{$m_2$}
\includegraphics[width=0.4\textwidth]{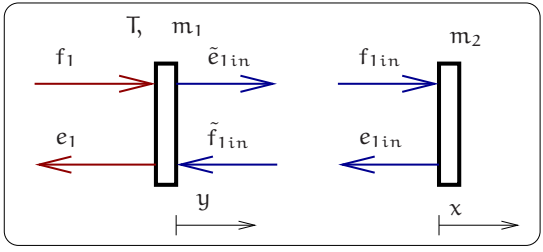}
\caption{Scheme of FP cavity  and notations.
        Both mirrors can move as free masses.
        }\label{FP}
\end{figure}

Here we derive formulas (\ref{f1in}, \ref{e1}) and also obtain the equations describing opto-elastic coupling in FP cavity with two movable mirrors (see notations in fig.~\ref{FP}).

We denote distance between FP cavities mirrors as $L$. Close to resonance case
it is convenient to introduce ``generalized transparency'' ${\cal T}_\Omega$ and
``generalized reflectivity'' ${\cal R}_\Omega$ for Stokes mode (\ref{TR}) using the following approximation:
\begin{align*}
\theta&= e^{i(\omega_1 +\Omega) \tau}\simeq 1+ i\Omega \tau,\quad
	\tau=\frac {L}{c}\, .
\end{align*}
Close to resonance we use the following approximation
$ \Omega  \tau\ll 1,\quad T\ll 1$.

The mean amplitude of main mode falling on end mirror is ${\cal F}_{1in}$ and we assume it as  a constant.

In general case  for fields on end mirror we have
\begin{align*}
&\sum_n\frac{{\cal A}_{1in}^{(n)}}{\sqrt {S_1^{(n)}}}\, e_{1in}^{(n)}e^{-i\omega_1 t}=
	- \sum_n\frac{{\cal A}_{1in}^{(n)}}{\sqrt {S_1^{(n)}}}f_{1in}^{(n)}e^{-i\omega_1 t}\, -\\
&\qquad -\frac{{\cal A}_{0in}}{\sqrt S_0}\, {\cal F}_{1in}e^{-i\omega_0t}2iku_\bot
	\left(xe^{-i\omega_m t}+x^*e^{i\omega_mt}\right).
\end{align*}
Here sum is taken over the complete set ${\cal A}_{1in}^{(n)}$ of cavity's modes (they are orthogonal to
each other) and $u_\bot$ is normal to surface component of displacement vector $\vec u$ of elastic mode
and $x$ is slow amplitude of displacement.
Multiplying this equation by distribution function ${\cal A}_{1in}^*$ of our Stokes mode, integrating over cross section and omitting non-resonance term ($\sim xe^{-i\omega_m t}$) one can find in frequency domain
\begin{align}
e_{1in}(\Omega)&=- f_{1in}(\Omega) - N_1 {\cal F}_{1in}2ik x^*(\Delta-\Omega) ,\\
\label{N1}
N_1 & = \frac{\int{\cal A}_{0in}{\cal A}_{1in}^*u_\bot\, d\vec r_\bot}{\sqrt {S_0S_1}}.
\end{align}

For fields on input mirror we have (see notations in fig.~\ref{FP})
\begin{align}
\label{fl2a}
\tilde e_{1in}&= iT\,f_1 - R\tilde f_{1in} +
        R( -{\cal F}_{1in})N_1\, 2ik y^*,\\
\label{fl2b}
e_1&= iT \tilde f_{1in} - R f_1  -RN_1{\cal F}_{1in}\, 2i\omega_1 y^*/c,\\
\label{fl2c}
f_{1in} & =\tilde e_{1in} \theta,\quad \tilde f_{1in}=\theta e_{1in}\, .
\end{align}

Combining these equations one can obtain:
\begin{align}
\label{b2r}
f_{1in}&={\cal T}_{\Omega}\, f_1+  \frac{N_1{\cal F}_{1in}{\cal T}_\Omega}{iT}\,2ik(x^*-y^*)
        ,\\
f_{1in} & \simeq \tilde e_{1in},\quad \tilde f_{1in}\simeq  e_{1in}=-\tilde e_{1in},\\
\label{2Fe}
e_1 &= f_1\,{\cal R}_{\Omega} - N_1{\cal F}_{1in}{\cal T}_\Omega\, 2ik(x^*-y^*).
\end{align}

One can find equation for amplitude $f_{1in}$ in time domain using the rule $(-i\Omega)\ \to\ \partial_t$:
\begin{align}
\label{f1inA}
\dot f_{1in} +\gamma f_{1in}= \frac{N_1{\cal F}_{1in}{\cal T}_\Omega}{iT}\,
	2ik(x^*(t)-y^*(t))e^{-i\Delta t}\, .
\end{align}

The light pressures acting on end mirror and input mirror inside the cavity are approximately equal to each
other. So we can calculate light pressure acting only on end mirror.
Keeping only cross term we can obtain formula for light pressure $P$ in time domain:
\begin{align*}
P&\simeq \frac{2 }{c\sqrt{ S_0S_1}} \left(
	{\cal A}_{0in}{\cal A}_{1in}^*{\cal	F}_{1in}f_{in1}^*(t)\,e^{-i(\omega_0-\omega_1)}+
	\right.\\&
\qquad \left. +{\cal A}_{0in}^*{\cal A}_{1in}{\cal F}_{1in}^*f_{in1}(t)\,e^{i(\omega_0-\omega_1)}\right).
\end{align*}
Now we can write down equation for elastic  oscillations with amplitude $x$:
\begin{align*}
\rho\sum_\ell   \vec u_\ell \big(\ddot x_\ell +2\gamma_m^{(\ell)} \dot x_\ell+(\omega_m^{(\ell)})^2 x_\ell\big)=
	\vec n_\bot P(\vec r_\bot)\, \delta(r_\| -r_\|^0)\, .
\end{align*}
Here $\rho$ is density of mirror,  sum is taken over the complete set of elastic modes (spatial displacement
vectors $ \vec u_\ell$ are orthogonal to each other), $\vec n_\bot$ is unit normal to mirror's surface, $r_\|$
is longitudinal coordinate of points inside body of mirror, coordinate $r_\|^0$ corresponds to face surface of
mirror. Multiplying this equation by spatial distribution vector $\vec u^*$ of our elastic mode
and integrating over mirror volume $V$ one can obtain:
\begin{align}
\ddot x +2\gamma_m \dot x+\omega_m^2 x & =\frac{2\left(
		N_1^*{\cal F}_{1in}^*f_{in1}(t)\,e^{i(\omega_0-\omega_1)t}+
		\text{c.c.}\right)}{cm\mu},\nonumber\\
\label{mu}
\mu & = \frac{1}{V}\int |\vec u(\vec r)|^2 \, d\vec r,
\end{align}
where $m=\rho V$ is mirror's mass.

Introducing the slow amplitudes $x\Rightarrow xe^{-i\omega_m t} +x^*e^{i\omega_m t} $  we  write down
 equation for amplitude $x^*$:
\begin{align}
\label{forceA}
\dot x^*+\gamma_m x^* &=
	\frac{N_1^*}{icm \mu\omega_m}
		{\cal F}_{1in}^*f_{in1}(t)\,e^{i\Delta t}\, .
\end{align}
And for the coordinate $z=x-y$ we finally obtain (if FP mirrors are elastically identical):
\begin{align}
\label{forceA2}
\dot z^*+\gamma_m z^* &=
	\frac{2N_1^*}{icm \mu\omega_m}
		{\cal F}_{1in}^*f_{in1}(t)\,e^{i\Delta t}\, .
\end{align}

\section{Symmetric mode}\label{PR}

In this Appendix we derive equations (\ref{fin+}, \ref{eqz+}) for analysis of parametric instability in symmetric mode.
One can start from equations for amplitudes $F_3$ and $E_3$ on beam splitter:
\begin{eqnarray}
F_3e^{-i\phi_{pr}} &=& i \sqrt{T_{pr}} F_5-\sqrt{1-T_{pr}} E_3 e^{i\phi_{pr}},\\
\label{E5}
E_5 &=&i\sqrt{T_{pr}} E_3 e^{i\phi_{pr}}- \sqrt{1-T_{pr}} F_5,\\
\phi_{pr} &=& \big(\omega_0+\Delta _{pr} +\Omega\big)l_{pr}/c.
\end{eqnarray}
We assume that PR cavity is in resonance:  $\exp(i\phi_{pr})= i $ and  we
assume that $\phi_{pr}$ does not depend on frequency $\Omega$ due to shortness of
PR cavity ($l_{pr}\ll L$). Then using (\ref{e3}) one can obtain:
\begin{align}
\label{f3b}
f_3  &\simeq \, \frac{\gamma-i\Omega}{( \gamma_{0+} - i\Omega)}
	\frac{-f_5\sqrt{T_{pr}}}{1+\sqrt{1-T_{pr}}}  -\\
&\qquad -	\frac{2\sqrt 2\,\gamma {\cal F}_{1in}\sqrt{1-T_{pr}}\,N_1
	kz_+^*}{\sqrt T
	\left(1+\sqrt{1-T_{pr}}\right) (\gamma_{0+} - i\Omega)},\nonumber\\
\label{e5}
e_5&=\frac{f_5\,(\gamma_{0+}+i\Omega)}{(\gamma_{0+}-i\Omega)} +
	\frac{\sqrt T_{pr} {\cal F}_{1in}\gamma\, 2\sqrt 2\,N_1 kz^*_+}{\sqrt T\,
		\big(1+\sqrt{1-T_{pr}}\big)(\gamma_{0+}-i\Omega)},\\
\label{gamma0+}
\gamma_{0+} &= \gamma\, \frac{1-\sqrt{1-T_{pr}}}{1+\sqrt{1-T_{pr}}}\simeq
	\frac{T_{pr}\gamma}{4}\, .
\end{align}

\paragraph*{The fluctuation part $f_+$ of symmetric mode.} We rewrite equation
(\ref{f1in}) using (\ref{f+}, \ref{f3b}):
\begin{align}
\label{fin+0}
f_{in+}  &=	 \frac{2i\sqrt{T_{pr}}\gamma\,f_{5} }{\sqrt{T}
	\big(1+\sqrt{1-T_{pr}}\big)(\gamma_{0+}-i\Omega)}+\\
&\quad +
	\frac{{\cal F}_{1in}\,2 \sqrt 2 i\,\gamma\, k N_1 z_+^* }{ T
	(\gamma_{0+}-i\Omega)}\, .\nonumber
\end{align}
Recall that in frequency domain  we mean
$z_+^*=z_+^*(\Delta -\Omega)$  ($\Delta $ is a detuning
(\ref{Deltaomega})).

Now we can write down Eq.~(\ref{fin+0}) in time domain using rule
$-i\Omega \, \to \, \partial_t$:
\begin{align}
(\partial_t+\gamma_{0+}) f_{in+}&=
	\frac{i\sqrt{T_{pr}}\gamma\,f_{5} }{\sqrt{T}}+
	\frac{2\sqrt 2 \, i{\cal F}_{1in}\,N_1 \gamma\, k z_+^* }{T}\, .
	\nonumber
\end{align}
For analysis of parametric instability in last equation we omit
term proportional to $f_5$ and finally obtain Eq.~(\ref{fin+}).

Then we can calculate the ponderomotive forces acting on each mirror and write equation for sum coordinate $z_+$:
\begin{align}
\label{Fpm+}
\frac{F_{pm+}}{m}&= \ddot z_+ +2\gamma_m \dot z_+
	+\omega_m^2 z_+,\\
\label{Fpm+2}
F_{pm+}&\simeq \frac{2\sqrt 2\, N_1}{\pi} \left(
	{\cal	F}_{1in}f_{in+}^*(t)\,e^{-i(\omega_0-\omega_1)}+	\right.\\&
\qquad \left. +{\cal F}_{1in}^*f_{in+}(t)\,e^{i(\omega_0-\omega_1)}\right).
	\nonumber
\end{align}
We rewrite (\ref{Fpm+2}) keeping only resonance terms:
\begin{align}
F_{pm+}&\simeq \frac{2\sqrt 2\, N_1}{\pi} {\cal F}_{1in}^*
	f_{in+}(t)\,e^{i(\omega_0-\omega_1)t},\\
\frac{F_{pm+}}{2i\omega_m m}&= e^{i\omega_m t}
	\left(\dot z_+^* +\gamma_m  z_+ ^*\right).
\end{align}
Now we can obtain equation (\ref{eqz+}).

\section{Anti-symmetric mode}\label{SR}

In this Appendix we derive equations (\ref{fin-}, \ref{eqz-}) for analysis of parametric instability in
anti-symmetric mode.
For amplitudes $F_4$ and $E_4$ on beam splitter we have:
\begin{eqnarray}
F_4e^{-i\phi} &=& i \sqrt{T_{sr}} F_6-\sqrt{1-T_{pr}} E_4 e^{i\phi},\\
\label{E6}
E_6 &=&i\sqrt{T_{sr}} E_4 e^{i\phi}- \sqrt{1-T_{sr}} F_6.
\end{eqnarray}
We assume that SR cavity is not in resonance
(i.e. $\phi=(\omega_0+\Omega)l_{sr}/c$ is an arbitrary number) and
that $\phi$ does not depend on frequency $\Omega$ due to shortness of SR
cavity ($l_{sr}\ll L$). Then using (\ref{e4}) one can obtain:
\begin{align}
\label{f4b}
f_4  &=\frac{if_6\sqrt{T_{sr}} e^{i\phi} (\gamma-i\Omega)}{
	\big(1+\sqrt{1-T_{sr}}\, e^{2i\phi}\big)\big(\gamma_{0-}-i(\delta+\Omega)\big)} +\\
&\quad +	\frac{2\sqrt2\,i\,  \gamma\,e^{2i\phi}\sqrt{1-T_{sr}}\,{\cal F}_{1in}N_1kz^*_-}{\sqrt T
	\big(1+\sqrt{1-T_{sr}}\, e^{2i\phi}\big)\big(\gamma_{0-}-i(\delta+\Omega)\big)},
	\nonumber\\
\label{e6b}
e_6 &= \frac{f_6\,e^{2i\phi}\,\big(\gamma_{0-}+i(\delta+\Omega)\big)}{
	 \big(\gamma_{0-}-i(\delta+\Omega)\big)} +\\
&\quad +\frac{2\sqrt2\,  \gamma\,\sqrt T_{sr}\, e^{i\phi}\,{\cal F}_{1in} \, N_1kz^*_-}{\sqrt T
	\big(1+\sqrt{1-T_{sr}}\, e^{2i\phi}\big)\big(\gamma_{0-}-i(\delta+\Omega)\big) }\, ,
	\nonumber\\
\label{delta}
\delta& =\frac{\gamma \sqrt{1-T_{sr}} \sin 2\phi}{\big(1-T_{sr}/2+\sqrt{1-T_{sr}}
	\cos 2\phi\big)},\\
\label{gamma0-}
\gamma_{0-} &=\frac{\gamma T_{sr}}{2\big(1-T_{sr}/2+\sqrt{1-T_{sr}} \cos 2\phi\big)}.
\end{align}

We rewrite equation (\ref{f1in}) using (\ref{f-}, \ref{f4b}) to obtain formula for fluctuation
part $f_-$ of anti-symmetric mode:
\begin{align}
\label{fin-0}
f_{in-}&= \frac{2\,\gamma \sqrt{T_{sr}}\,f_{6} e^{i\phi}}{\sqrt{T}
	\big(1+\sqrt{1-T_{sr}}\, e^{2i\phi}\big)
	\big(\gamma_{0-}-i(\delta+\Omega)\big)} +\\
&\qquad +\frac{2\sqrt 2\,i\, \gamma \,{\cal F}_{1in}\, N_1k
	 z^*_-}{T\big(\gamma_{0-}-i(\delta+\Omega)\big)}.\nonumber
\end{align}

For analysis of parametric instability we can omit the term proportional to $f_6$ so that
\begin{align}
\label{finO-}
f_{in-}&=\frac{N_1 {\cal F}_{1in}\,i\, \omega_1z^*_-}{L\sqrt 2
	\big(\gamma_{0-}-i(\delta+\Omega)\big)}\, .
\end{align}
Finally one can obtain from (\ref{finO-}) the equation (\ref{fin-}) in time domain.

In the same manner as for symmetric mode one can obtain the formula for ponderomotive force in
anti-symmetric mode:
\begin{align}
\label{Fpm-}\frac{F_{pm-}}{m}&= \ddot z_- +2\gamma_m \dot z_-
	+\omega_m^2 z_-,\\
\label{Fpm-2}
F_{pm-}&\simeq \frac{2\sqrt 2\, S}{\pi} \left(
	{\cal	F}_{1in}f_{in-}^*(t)\,e^{-i(\omega_0-\omega_1)t}+	\right.\\&
\qquad \left. +{\cal F}_{1in}^*f_{in-}(t)\,e^{i(\omega_0-\omega_1)t}\right).
	\nonumber
\end{align}
Keeping only resonance terms and rewriting  (\ref{Fpm-},\ref{Fpm-2}) one can
obtain equation (\ref{eqz-}) for differential coordinate $z_-$ in time domain.

\section{Only one mirror is in resonance}\label{PRSR}

For ponderomotive force we have
\begin{align}
F_{pm}&=\frac{2S(E_\text{main}+E_\text{Stokes})^2}{4\pi}\simeq\\
&\simeq \frac{2}{c\mu}\left({N_1\cal F}_{1in}(t)\,f_{in1}^*(t) +
	N_1^*{\cal F}_{1in}^*(t)\,f_{in1}(t)\right)\, .
	\nonumber
\end{align}
Keeping only resonance term ($\sim f_{in1}$) we obtain equation for elastic mode:
\begin{align}
\dot x_1^* +\gamma_m x_1^* &= \frac{N_1^*{\cal F}_{1in}^*e^{i\Delta t}}{ ic\mu \, m\omega_m}
	f_{1in}(t),\quad
	\Delta =\omega_0-\omega_1 -\omega_m\, .\nonumber
\end{align}

Taking into account both  anti-symmetric  and symmetric modes  we finally obtain the set of equations in time domain:
\begin{align}
\label{x1a}
\dot x_1^* +\gamma_m x_1^* - \frac{N_1^*{\cal F}_{1in}^*}{ic\mu \,m \omega_m}
	\frac{f_{in+}+f_{in-}}{\sqrt 2}e^{i\Delta t}&=0\, ,\\
\label{fina+}
-	\frac{ iN_1{\cal F}_{1in}\,  \omega_1 }{\sqrt 2\, L} \,x_1^*e^{-i\Delta  t}+
	(\partial_t +\gamma_{0+}) f_{in+} &=0\, ,\\
\label{fina-}
-	\frac{ iN_1{\cal F}_{1in}\,  \omega_1 }{\sqrt 2\, L}
	\,x_1^*e^{-i\Delta  t}+(\partial_t -i\delta+\gamma_{0-}) f_{in-} &=0\, .
\end{align}
We sum and subtract equations (\ref{fina+}, \ref{fina-}) (see also definitions (\ref{f+}, \ref{f-})):
\begin{align}
\label{f1x}
\dot f_{1in}+ \frac{\gamma_{0+}+\Gamma_{0-}}{2}\,f_{1in} +
	\frac{\gamma_{0+}-\Gamma_{0-}}{2}\,f_{2in} &=\sigma\,  ,\\
\label{f2x}
\dot f_{2in}+ \frac{\gamma_{0+}+\Gamma_{0-}}{2}\,f_{2in} +
	\frac{\gamma_{0+}-\Gamma_{0-}}{2}\,f_{1in} &= 0\, ,
\end{align}
where we introduced notations:
\begin{align}
\sigma&\equiv \frac{ iN_1{\cal F}_{1in}\,  \omega_1 }{ L}\,x_1^*e^{-i\Delta  t},
	\quad  \Gamma_{0-}\equiv \gamma_{0-}-i\delta \,.  \nonumber
\end{align}
Manipulating Eqs. (\ref{f1x}, \ref{f2x}) we get:
\begin{align}
\left (\partial_t + \frac{\gamma_{0+}+\Gamma_{0-}}{2}\right)\times (\ref{f1x})
	- \left(\frac{\gamma_{0+}-\Gamma_{0-}}{2}\right)\times (\ref{f2x})\, :\nonumber\\
\ddot f_{1in} +(\gamma_{0+}+\Gamma_{0-})\dot f_{1in}+
	\gamma_{0+}\Gamma_{0-}\, f_{1in} =
		\dot \sigma + \frac{\gamma_{0+}+\Gamma_{0-}}{2}\, \sigma\, .\nonumber
\end{align}
Finally we have the set of equations:
\begin{align}
\label{x1b}
	\dot x_1^* +\gamma_m x_1^* - \frac{N_1^*{\cal F}_{1in}^*}{ic\mu\, m\omega_m}
	f_{1in}(t)e^{i\Delta t}&=0\, ,\\
\frac{- iN_1{\cal F}_{1in}\,  \omega_1 }{ L}
	\left(\partial_t + \frac{\gamma_{0+}+\Gamma_{0-}}{2}\right)x_1^*e^{-i\Delta t}+&\\
\quad	+\ddot f_{1in} +(\gamma_{0+}+\Gamma_{0-})\dot f_{1in}+
	\gamma_{0+}\Gamma_{0-}\, f_{1in} &=0\, . \nonumber
\end{align}
Finding the solution of this set in form: $f_{1in}(t)=f_{1in}e^{\lambda t }, \quad x_1^*(t)= x_1^*e^{(\lambda+i\Delta) t }$ we obtain characteristic equation:
\begin{align}
\frac{\big(\lambda+i\Delta +\gamma_m\big)
		\big(\lambda + 	\gamma_{0+}\big)\big(\lambda+\Gamma_{0-}\big)}{
		\left(2\lambda +\gamma_{0+}+\Gamma_{0-}\right)}=&
		\frac{\cal Q}{2}\, .\nonumber\
\end{align}
Using this equation we get Eq. (\ref{CharEqx1}).

\paragraph*{The instability condition  for pure SR configuration.}
As before we substitute  $\lambda=a +i(b-\Delta)$ into (\ref{CharEqx1}) taking $a=0$:

\begin{align}
\label{SRg1}
\gamma_m&=
  \underbrace{\gamma_{0+}A_1}_{\gamma_{m1}}+\underbrace{\gamma_{0-}A_2}_{\gamma_{m2}},\\
\label{SRb2}
  b&= A_1(\Delta-b)+A_2(\Delta+  \delta -b)\, .
\end{align}
Here we introduce notations
\begin{align}
A_1&=\frac{{\cal Q}}{2}
  \frac{ 1}{\gamma_{0+}^2+(b-\Delta)^2},\quad \gamma_{m1}=\gamma_{0+}A_1,\\
 A_2&=\frac{\cal Q}{2}
  \frac{ 1}{\gamma_{0-}^2+(b-\Delta-\delta)^2},\quad \gamma_{m2}=\gamma_{0-}A_2\, .
\end{align}
 We can  formally solve Eq. (\ref{SRb2}) and find $b$:
\begin{align}
\label{bSR}
b=\frac{\dfrac{\gamma_{m1}}{\gamma_{0+}}\Delta+
	\dfrac{\gamma_{m2}}{\gamma_{0-}}(\Delta+\delta)}{1+
	\dfrac{\gamma_{m1}}{\gamma_{0+}}+\dfrac{\gamma_{m2}}{\gamma_{0-}}}\, .
\end{align}
Note that values $\gamma_{m1}$ and $\gamma_{m2}$ in (\ref{SRg1}) are positive ones and hence
$\gamma_{m1},\gamma_{m2}<\gamma_m$. Also taking into account that
$\gamma_{m1},\gamma_{m2}\ll\gamma_{0+},\gamma_{0-}$ we can
conclude from (\ref{bSR}) that
\begin{equation}
\label{smallb}
b\ll|\Delta|,\ |\Delta +\delta|\,.
\end{equation}
Then the
parametric instability condition (\ref{SRPIcond}) can be easily obtained from
Eq.~(\ref{SRg1}) using inequality (\ref{smallb}).

\section{Numerical Parameters}\label{param}

We used the parameters planned for Advanced LIGO \cite{ligo2}:
\begin{equation*}
\begin{array}{rclrcl}
\omega_0\simeq \omega_1 & \simeq & 2\times 10^{15}\ {\rm sec}^{-1},&  L &=&4\times 10^5\ {\rm cm},\\
m &=& 40\ \text{kg},  & W &=& 830\ {\rm kW},\\
T &=& 5\times 10^{-3}, &  T_{pr}&= & 6\times 10^{-2},\\
\gamma & \simeq & 94\ {\rm sec}^{-1},&
    \gamma_{0+} &\simeq  & 1.5\ {\rm sec}^{-1},\\
T_{sr} &=& 7\times 10^{-2}.
\end{array}
\end{equation*}

We also assume that  FP mirrors are fabricated from fused silica with angle of structural losses
$\phi= 1.2\times 10^{-8}$ and
for elastic modes frequencies  $\omega_m=10^5 \ \div \  10^7\ {\rm sec}^{-1}$ we estimate
relaxation rate $\gamma_m$ of elastic modes by formula:
$$
\gamma_m=\omega_m \phi/2
\simeq 6\times \big( 10^{-4}\ \div \ 10^{-2})  {\rm sec}^{-1}.
$$

\end{document}